\documentclass[prl,twocolumn,preprintnumbers,amsmath,amssymb,superscriptaddress]{revtex4}
\usepackage{graphicx}
\usepackage[ansinew]{inputenc}
\usepackage{array}
\usepackage{color}
\usepackage{amsmath}
\usepackage{amsxtra}
\usepackage{amstext}
\usepackage{amssymb}
\usepackage{latexsym}

\newcommand\revision[1]{\textcolor{black}{#1}}

\begin{document}

\title{Metamorphosis of Goldstone and Soft Fluctuation Modes in Polariton Lasers}

\author{R. Binder}
\affiliation{Wyant College of Optical Sciences, University of Arizona, Tucson, AZ 85721}
\affiliation{Department of Physics, University of Arizona, Tucson, AZ 85721}

\author{N.H. Kwong}
\affiliation{Wyant College of Optical Sciences, University of Arizona, Tucson, AZ 85721}


\date{\today}

\begin{abstract}
For a driven-dissipative quantum many-body system prepared in a spontaneous broken-symmetry
steady state, in addition to the Goldstone mode  the soft fluctuation modes provide important
insight into the system's dynamics.  Using a microscopic polariton laser theory, we find
a rich transformation behavior of discrete and continuum soft modes in a two parameter
(pump density and cavity dissipation rate) space.
\revision{
In addition to finding exceptional-point Goldstone companion modes,
our theory yields a unified
picture of a variety of seemingly disconnected physical concepts including Mott transition,
Mollow spectra
or relaxation oscillations, and
polaritonic Bardeen-Cooper-Schrieffer gaps.
}
\end{abstract}

\maketitle


Electrons, holes, and photons in an excited semiconductor microcavity
have been studied extensively (e.g.\
\cite{%
fan-etal.97pra,%
cao-etal.97,%
gonokami-etal.97,%
kira-etal.99b,%
moskalenko-snoke.00,%
Ciuti2000,%
Savvidis2000,%
kwong-etal.01prl,%
Baumberg2005,%
Keeling2007,%
schumacher-etal.07prb,%
bajoni-etal.08,%
berney-etal.08,%
amo-etal.09,%
sanvitto-timofeev.12,%
semkat-etal.09,%
kamide-ogawa.10,%
deng-etal.10,%
snoke-littlewood.10,%
liu-etal.14,%
schulze-etal.14,%
menard-etal.14,%
kamandar-etal.14,%
schmutzler-etal.15,%
leeuw-etal.16,%
hayenga-etal.17,%
kavokin-etal.17,%
bao-etal.19}).
They form a quantum many-body system that can organize itself into configurations that
support coherent order parameters.
A semiconductor microcavity laser is a prime example.
Being a many-body system with strong photon-matter and Coulomb couplings, it can undergo the lasing transition through more than one physical mechanisms. While the `conventional' mechanism, with the photons as the main coherent field, is valid for many semiconductor lasers, in some cases lasing is found to be the result of a transition to a state in which polaritons form a nonequilibrium analog of a Bose-Einstein (BEC) or BCS condensate
\cite{szymanska-etal.06,deng-etal.10,hu-etal.2019arxiv}.
Much insight on these broken-symmetry states can be gained by probing their fluctuation modes. Of particular interest is the Goldstone mode, which is
\revision{a phase mode of the coherent laser field},
 and the dissipative modes that are associated with it. It was shown in previous theoretical works on
atomic lasers \cite{chou-su.86} and pumped-dissipative exciton-polariton condensates in quantum-well microcavities \cite{szymanska-etal.06,wouters-carusotto.07a,hanai-etal.18}
that at zero momentum, the Goldstone mode
is accompanied by a damped mode that shares its frequency
\revision{
(here called Goldstone companion mode),
}
and the dispersion
\revision{ of the Goldstone mode }
  at low momentum is diffusive \cite{szymanska-etal.06,wouters-carusotto.07a}. Absent in equilibrium BEC or BCS, this second mode is
characteristic of the nonequilibrium character of the laser. It coincides with the Goldstone mode at the lasing threshold and acquires a growing decay rate with further (small) increase in the lasing intensity.
Since the laser is a spatially extended many-particle system, it should be expected that the two modes, Goldstone and companion, are in the vicinity, energy-wise, of a large (possibly infinite) number of
\revision{
fluctuation modes with discrete energies or in spectral continua.
}
However, not much is known about this complex landscape of modes and how they transform as external parameters are varied.
As we show below, our theoretical analysis provides a unified picture for the transformation of discrete and continuous sets of linear excitation modes, from
below threshold (polariton modes and electron-hole continua) to above threshold (Goldstone and companion modes), thus clarifying the relation of
seemingly disconnected physical concepts including Mott transition \cite{semkat-etal.09},
as well as
light-induced gaps \cite{galitskii-etal.70},
Mollow-like spectra  and side-band emission \cite{quochi-etal.98,berney-etal.08,byrnes-etal.10,horikiri-etal.16},
 polaritonic BCS gaps \cite{szymanska-etal.06,hu-etal.2019arxiv}, and Goldstone modes \cite{wouters-carusotto.07a,hanai-etal.18}.
 \revision{
 Relaxation oscillations in semiconductor lasers \cite{chow-etal.94,keller-gallmann}   have been interpreted as
 dynamical Stark effect (Mollow) sidebands in \cite{thedrez-frey.88}, and
 relaxation oscillations in polariton condensates have been observed \cite{giorgi-etal.14,horikiri-etal.18}.
 }
 Furthermore,
 \revision{ we find that }
 above threshold the mode transformations include exceptional points \cite{heiss.04}.

We use a semiclassical microscopic theory detailed in the Supplementary Material, with electrons, holes, and photons as degrees of freedom, to analyse an incoherently pumped
 GaAs quantum well inside a microcavity, and to calculate the steady state configurations and their linear response to an interband optical probe. The fluctuation modes underlying the response are obtained by formulating the response as an eigenvalue problem  which is solved numerically. The probe is normal to the quantum well's plane, and hence the probed fluctuations are those of
 \revision{zero wave vector $\textbf{q}$ in the plane (we do not consider the dispersion of the modes as a function of $\textbf{q}$}).
 The eigenvalue set generally consists of continua and discrete modes distributed on the complex energy plane.
We follow the evolution of the eigenvalues as we vary the pump density $n_{\rm p}$ and the cavity loss rate $\gamma_{\rm cav}$. The dissipative Goldstone companion mode \cite{szymanska-etal.06,wouters-carusotto.07a,hanai-etal.18}, alluded to above, is seen to undergo some non-trivial transformation in this parameter space. At sufficiently high $n_{\rm p}$, and sufficiently large $\gamma_{\rm cav}$, it meets another laser-frequency, dissipative mode at an exceptional point
\cite{%
heiss.04,%
dembowski-etal.01,%
liertzer-etal.12,%
gao-etal.15,%
hanai-etal.19,%
hanai-littlewood.19,%
miri-etal.19,%
sakhdari-etal.19%
},
where the two modes turn into two others of finite frequency. These latter modes evolve upon further parameter change to Mollow-like modes.

\begin{figure}[t]
\includegraphics[width=.5 \textwidth]{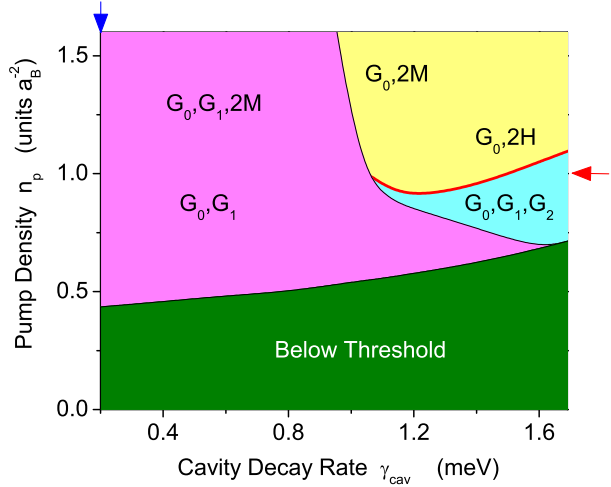} 
\caption{ (Color Online)
A
\revision{ map }
of the discrete modes. Each region is marked by the modes present.
Changes in the set of modes occur at the regional boundaries. ${\rm G}_0$: Goldstone mode,
${\rm G}_1$,${\rm G}_2$: damped modes at the same frequency as ${\rm G}_0$, H,M: Mollow analog modes.
The meaning of ${\rm G}_1$, ${\rm G}_2$, H and M  will be clarified in the following figures. Each point
on the red curve separating the blue and yellow regions is an exceptional point.
}
\label{fig:sr1600etcSchematic}
\end{figure}

 To
provide a unifying picture for this and other interesting relations among the subset of discrete soft modes, we show in Fig.
\ref{fig:sr1600etcSchematic} a
\revision{ map }
of the discrete modes. The details of this
figure and also the role of continua of modes
will become more clear throughout this paper. The response spectra governed by the fluctuation modes are also calculated and shown below.
We find the
spectral features to be generally consistent with experimental photoluminescence data
found in the literature \cite{quochi-etal.98,horikiri-etal.16}. Our fluctuation analysis provides detailed understanding of these spectra and identifies parameter regimes where new interesting effects may be observed.

Our theory is based on the semiconductor Bloch equations (SBE) \cite{chow-etal.94}
 amended by a
single-mode equation for the light field in the cavity $E$. The equations
are of the form $i\hbar \mathbf{\dot{A}=F(A)}$ where the dynamical variables
$\mathbf{A}=(P(\mathbf{k}),f(\mathbf{k}),E)^{T}$ contain, in addition to $E$,
the complex--valued interband polarization $P(\mathbf{k})$ and real--valued
carrier distribution $f(\mathbf{k})$ as a function of wave vector $%
\mathbf{k}$, and $\mathbf{F(A)}$ is a nonlinear function that contains
Coulomb interaction, relaxation, decay \ and source (incoherent pump)
terms. We solve the equations numerially until steady state is reached. The
steady state solution, $\mathbf{A}^{(0)}=(P^{(0)}(\mathbf{k}),f^{(0)}(%
\mathbf{k}),E^{(0)})^{T}$ is non-zero above threshold. A small external
perturbation field $E_{pert}$ can  probe the steady state by inducing small
changes $\mathbf{x}=(\delta P(k),\delta P^{\ast }(k),\delta f(k),\delta
E,\delta E^{\ast })^{T}$. Linearizing the function $\mathbf{F(A}^{(0)}+%
\mathbf{x)}$ then yields an equation of the form $i\hbar \mathbf{\dot{x}\ =}M%
\mathbf{x}+\mathbf{s}_{pert}$ where the complex-valued non-symmetric matrix $%
M$ is a nonlinear function of $\mathbf{A}^{(0)}$, and $\mathbf{s}_{pert}$ is
proportional to $E_{pert}$. The right-eigenvalue equation is  $M\mathbf{x}%
^{(n)}\mathbf{=\varepsilon }^{(n)}\mathbf{x}^{(n)}$.
In all the figures displaying eigenvalues, the zero of the real frequency axis is
set at the lasing frequency when the system is above the lasing threshold and at the
lower polariton (LP) frequency below threshold. We refer to all the discrete modes
that oscillate at the laser frequency collectively as G modes, and label them by
${\rm G}_n , n = 0,1,2,...$, with ${\rm G}_0$ being the Goldstone mode. These modes
have zero real parts in our plots.
We define the
complex-valued linear-response function $\chi (\omega )=\delta
P_{tot}(\omega )/E_{pert}(\omega )$ where $\delta P_{tot}(\omega )\propto
\sum_{k}\delta P(k,\omega )$.

\begin{figure}[t]
\includegraphics[width=.5 \textwidth]{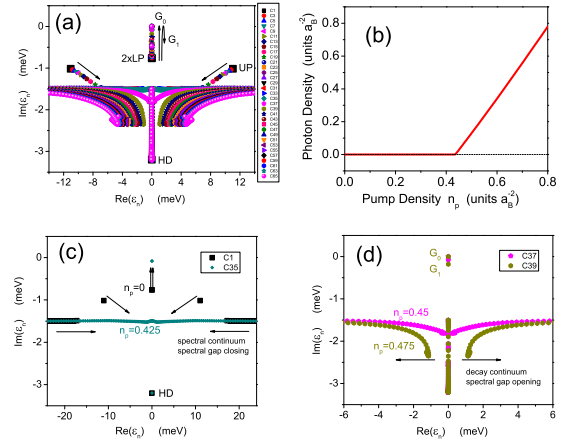} 
\caption{ (Color Online)
(a) Eigenenergies obtained from the diagonalization of the linear response matrix $\hat{M}$ for equidistantly spaced pump densities
from almost  zero, $n_p = 1\times 10^{-4} a_B^{-2}$,   (C1)  to $0.8 a_B^{-2}$ (C65),
 at T=50K and cavity decay $\gamma_{cav} = 0.2$meV.
 Number of k-points is $N_k = 300$, linear dimension of  $\hat{M}$ is $3N_k + 2$.
At zero density we see a damped 2-fold degenerate
lower polariton
(LP) state, two upper polariton (UP) states (symmetric above and below LP), and
a highly-degenerate (HD) damped state, with a degeneracy of $N_k$.
The arrows indicate the evolution of eigenvalues with increasing pump density.
(b) Input-output curve (photon density vs. pump density), with threshold approximately at 0.435 $a_B^{-2}$.
(c) Same as (a) but only showing for two pump densities (almost zero and just below threshold). Spectral gap closing through merging of spectral continua is indicated.
(d) Same as (c) for two pump densities just below (above) generation of decay continuum and (BCS-like) spectral gap opening.
Here and in all figures showing eigenenergies $\varepsilon_n$, the zero of the real axis is at the LP (below threshold) or the laser frequency (above threshold).
The zero of the imaginary axis separates decay (Im$\varepsilon_n < 0$) from growth (Im$\varepsilon_n>0$).
}
\label{fig:sr1611zeigval}
\end{figure}

\begin{figure}[t]
\includegraphics[width=.5 \textwidth]{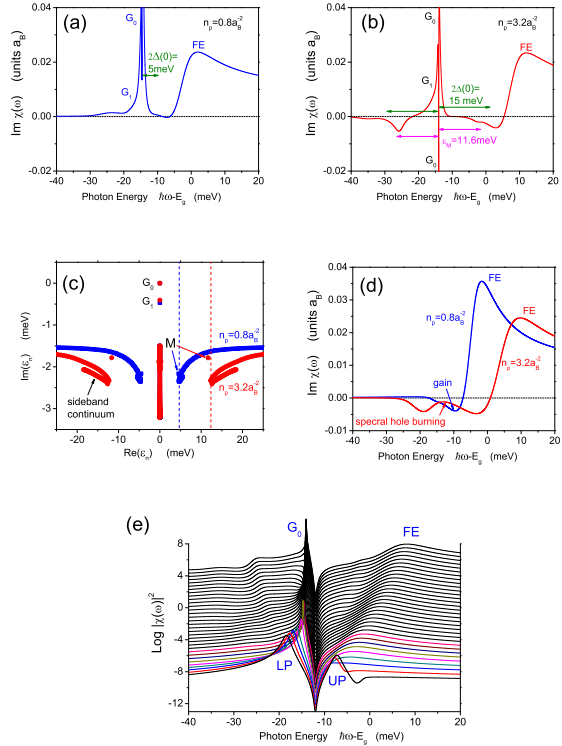} 
\caption{ (Color Online)
(a,b) Calculated linear response spectra, Im$\chi(\omega)$,  for two different pump densities
(above and far above threshold). ${\rm G}_0$ and ${\rm G}_1$ resonances and peak due to Fermi edge (FE) absorption are indicated.
BCS-like gap $2 \Delta ( {\bf 0} )$ (value taken from numerical solution)
and Mollow-like states $\varepsilon_M$ (value taken from part (c) of this figure) are indicated.
(c) Eigenenergies corresponding to spectra in (a,b).
 The vertical dashed lines indicate the estimate of the pair excitation gap $\tilde{E}_{gap}^{pair}$.
(d) For comparison, the spectra obtained from conventional gain-spectra calculations, using same distribution functions
as in (a,b) but without coherent light and interband polarization fields.
(e) Calculated linear response spectra, $|\chi(\omega)|^2$,
 for pump densities equidistantly spaced from  almost  zero      to $n_p=3.2 a_B^{-2}$. Except for the lowest pump density,
the spectra are shifted vertically for clarity.
For clarity, the Fourier transform from time to frequency contains
 an additional phenomenological broadening $\gamma_{a} = 0.04$meV, since the Goldstone mode would diverge otherwise.
}
\label{fig:sr1613spectra}
\end{figure}

Fig. \ref{fig:sr1611zeigval} illustrates the evolution of the eigenvalue
set as a function of pump density, from near zero to $0.8 a_B^{-2}$, for a
fixed cavity decay rate of $\gamma_{cav}=0.2$meV. Part (a) gives an overview.
At vanishing density (C1 in the figure), the eigenvalue set consists of a
two-fold degenerate lower polariton (LP) mode and a highly degenerate (HD) mode
on the imaginary axis, two upper polariton (UP) modes and two spectral (along
the real energy direction) continua. The spectral continua show a gap at low frequency
due to exciton binding. The modes with non-zero frequency are symmetrically
placed. When the system is below threshold, the positive-energy
modes are associated with
$\delta P, \delta E$, and the negative-energy modes with their complex conjugates.
As the pump density $n_{\rm p}$ increases,
the LP modes move up towards zero damping, and the continuum gap closes, ionizing the UP
in Mott transition. As the density crosses the threshold, shown in part (b) to be at
$n_{\rm p} \approx 0.435 a_B^{-2}$,
 the LP modes transforms \cite{szymanska-etal.06,wouters-carusotto.07a} to the
Goldstone mode ${\rm G}_0$ and its damped counterpart
${\rm G}_1$, the continuum spectral gap closes completely, and the degenerate HD mode
spreads into a decay continuum (along the imaginary energy direction). As the density
increases further, the lasing field forces a reopening of the spectral gap via a mechanism
similar to that in BCS theory. These trends are shown in more detail in parts (c) and (d).
\revision{The BCS gap will be further clarified in Fig. \ref{fig:sr1613spectra}.
Regarding the physical nature of the modes, HD is a pure non-radiative density fluctuation, the decay
continuum modes are radiative (involving density, polarization and light field fluctuations),
the spectral continua do not contain density oscillations and their polarization oscillations
are sharply peaked as a function of $k$ (similar to ionization continuum wave functions of excitons or hydrogen),
$G_0$ ($G_1$) is a collective mode with a smooth variation of the polarization fluctuation as a function of $k$
and without (with) density fluctuation.}

Fig. \ref{fig:sr1613spectra}(a-d) show the spectral features that emerge at
higher pump densities. Parts (a) and (b) compare the response spectra
${\rm Im} \chi ( \omega )$ at $n_{\rm p} = 0.8 a_B^{-2}$,
which is the upper limit in Fig. \ref{fig:sr1611zeigval}, and
$n_{\rm p} = 3.2 a_B^{-2}$. $\gamma_{\rm cav}$ has the same value as in Fig.
\ref{fig:sr1611zeigval}. The modes underlying the features in the response spectra
are shown in Fig. \ref{fig:sr1613spectra}c. A discrete mode, in addition to ${\rm G}_0$
and ${\rm G}_1$, emerges and becomes separated from the spectral continuum
\revision{(a discrete mode at the edge of a continuum, rather than inside the continuum,
could be called semi-Fano resonance)}.
 We label
this mode M and identify it as an analog to the Mollow sideband modes
(cf.\ \cite{thedrez-frey.88,quochi-etal.98,berney-etal.08,byrnes-etal.10,horikiri-etal.18}).
 To support this interpretation, we note that the real part of
M's energy, denoted by $\varepsilon_{\rm M}$, is comparable in value to twice the effective
Rabi energy
in the rotating-wave approximation
$\Delta ( {\bf k} ) = a_{c}E+\sum\limits_{k^{\prime}}V(k-k^{\prime })
P(k^{\prime })$ (where $V(k-k^{\prime })$ is the Coulomb interaction and $a_{c}$ a coupling
constant) (see Eq. (15) in the Supplementary Material).
For example, for $n_{\rm p} = 3.2 a_B^{-2}$,
$\varepsilon_{\rm M} = 11.6 {\rm meV}$ and $2 \Delta ( {\bf k} = {\bf 0} ) = 15 {\rm meV}$.
\revision{
Note that $\sum\limits_{k^{\prime}}V(k-k^{\prime })$ is the standard BCS gap function, and
 $\Delta ( {\bf k} )$ can be viewed as a composite gap function for polariton systems, see, e.g.,
 \cite{yamaguchi-etal.13}.
}
The BCS-like spectral gap grows with $n_{\rm p}$ and its numerical
value, marked by the vertical dashed lines, agrees very well with an estimate of the
pair-breaking energy $\tilde{E}_{gap}^{pair}$, which is explained (Eq. (34)) in
the Supplementary Material. The lasing field couples the $\delta P, \delta E$ components
of the perturbed polariton field to their complex conjugates in a manner similar to four-wave
mixing (this coupling is absent in linear response below the lasing threshold).
\revision{The resulting sideband continuum}
is, as shown
in Fig. \ref{fig:sr1613spectra}c, much more pronounced at high $n_{\rm p}$. In parts (a) and (b),
the central peaks show the sharp spikes of the Goldstone mode ${\rm G}_0$, regularized by a small
damping width, and the broader contribution from ${\rm G}_1$. To assess the effect of the M modes,
we mark the magnitudes of $\varepsilon_{\rm M}$ and $2 \Delta ( {\bf 0})$ on the plots. While it is
not visible at $n_{\rm p} = 0.8 a_B^{-2}$, the M modes appear to produce features reminiscent of Mollow
sidebands in atomic physics at the higher density. This density, $n_{\rm p} = 3.2 a_B^{-2}$ may
however be too high for  experimental verification with stationary lasers.
For comparison, we also show
in Fig. \ref{fig:sr1613spectra}d
gain spectra
calculated
for an isolated quantum well, but with the
distribution functions obtained from the full calculation
\revision{
in which all G-modes are missing (cf. \cite{yamaguchi-etal.13,yamaguchi-etal.15}).
}
Far above
threshold, they develop signatures of spectral hole burning.
In Fig. \ref{fig:sr1613spectra}e we show a sequence of spectra $|\chi (\omega )|^{2}$, which is everywhere positive (as are photoluminescence spectra) and allows
us to follow the position of the various resonances over a large variation of
pump densities. This shows how the Fermi edge absorption grows out of the
UP, which was conjectured in \cite{quochi-etal.98}, and how the blue shift of the
Fermi edge absorption levels off at high densities, consistent with the
experimental findings in \cite{horikiri-etal.16}.

\begin{figure}[t]
\includegraphics[width=.5 \textwidth]{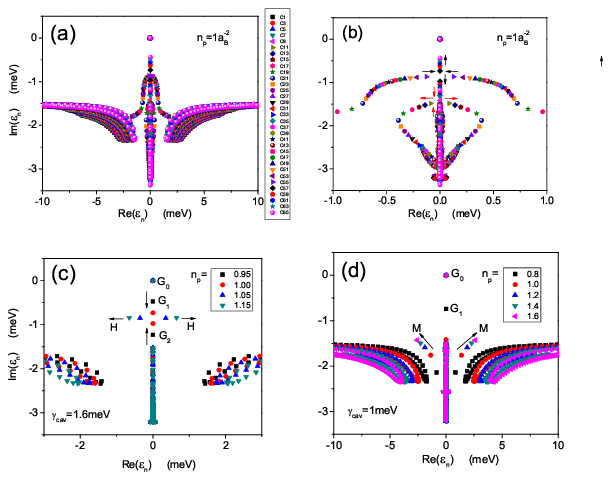} 
\caption{ (Color Online)
(a) Eigenenergies for  equidistantly spaced cavity rates $\gamma_{cav}$ from
  0.9meV (C1)    to 1.7 meV (C65) at fixed pump density
  $n_p=1 a_B^{-2}$.
  (b) Same as (a) but zoomed in to smaller frequency interval.
  The arrows indicate the evolution of eigenvalues with increasing decay rates.
  The transformation marked by the red arrows happens at a lower $\gamma_{\rm cav}$
  than that marked by the black arrows.
  In the transformation indicated by the black arrows, the two converging modes meet at an exceptional point.
  (c) Detail of a `collision', as $n_{\rm p}$ increases, of two damped G modes (${\rm G}_2$, ${\rm G}_3$) at an exceptional point generating two H-modes.
  (d) Example of Mollow-like modes (M) rising above decay continuum.
}
\label{fig:sr1600-37-59-zeigval}
\end{figure}

Fig. \ref{fig:sr1600-37-59-zeigval}(a), and (b) in more detail, show
the evolution of the eigenvalues for a fixed pump density $n_{\rm p} = 1 a_B^{-2}$
(indicated by the red arrow in Fig. \ref{fig:sr1600etcSchematic}) as the cavity decay
varies.
As $\gamma_{\rm cav}$ increases, ${\rm G}_1$ moves
down and merges with the decay continuum at $\gamma_{\rm cav} \approx 1.1$meV.
At this point, two modes come out of the continuum acquiring finite frequency.
This progression is indicated by the red arrows in part (b).
Upon further rise in $\gamma_{\rm cav}$, the Mollow analogs M modes cross over to low frequencies,
where we rename the modes as H to indicate that the modes' characteristics have changed.
As indicated by the black arrows in part (b), the two H modes meet at an exceptional point
\cite{%
heiss.04,%
dembowski-etal.01,%
liertzer-etal.12,%
gao-etal.15,%
hanai-etal.19,%
hanai-littlewood.19,%
miri-etal.19,%
sakhdari-etal.19%
},
where the two eigenvectors become one state which is self-orthogonal with respect to
the inner product using right and left eigenvectors, and turn into two G modes, ${\rm G}_1$
moving up the imaginary axis and ${\rm G}_2$ moving down. The bottom panels of
Fig. \ref{fig:sr1600-37-59-zeigval} show the generation
of the H/M modes as density increases by a transformation from ${\rm G}_1$ and ${\rm G}_2$
at an exceptional point (part (c)) and by separating from the spectral continuum (\revision{part} (d)).

As noted above, we have performed the eigenvalue computation over a region of the $( \gamma_{\rm cav} , n_{\rm p} )$ parameter plane and constructed
\revision{ a map}
of discrete fluctuation modes shown in Fig. \ref{fig:sr1600etcSchematic}.
\revision{
The lines separate regions of different numbers of discrete G modes
(unlike lines in phase diagrams that separate regions of different numbers of stable and unstable solutions to a nonlinear set of equations).
}
The specific cases discussed above are instances on this diagram: the blue arrow marks the cases in Figs. \ref{fig:sr1611zeigval} and \ref{fig:sr1613spectra}(a-c) and the red arrow marks the cases in Fig. \ref{fig:sr1600-37-59-zeigval}(a-b). The three (color-coded) regions above the lasing threshold are defined and labeled by the Goldstone and discrete soft modes that are present, and some modes are transformed when a curve between two regions is crossed. The (red) curve separating the blue and yellow regions is made up of exceptional points at which ${\rm G}_1$ and ${\rm G}_2$ meet and turn into two H modes or vice versa, as illustrated in Fig. \ref{fig:sr1600-37-59-zeigval}(a) and (b) and in detail in (c). In the pink region, as $n_{\rm p}$ increases from the lasing threshold, ${\rm G}_1$ separates from the Goldstone mode ${\rm G}_0$ and moves down along the imaginary axis in the complex energy plane. When the boundary curve with the blue region is crossed, the mode ${\rm G}_2$ emerges from the decay continuum and moves up towards ${\rm G}_1$ along this axis. ${\rm G}_2$ does not appear at small $\gamma_{\rm cav}$, and at the curve marking the crossing from the pink region to the yellow region, ${\rm G}_1$ merges with the decay continuum and ceases to be a distinct mode.

In summary, we have studied in detail the low-frequency fluctuation modes of an incoherently-pumped polariton laser at the long-wavelength limit. The set of mode eigenvalues generally consists of both continuous and discrete subsets in the complex energy plane. Tracing the evolution of this eigenvalue set when two parameters (pump density and cavity loss rate) vary, reveals interesting features. The onset of lasing spreads a highly degenerate dissipative mode into a decay continuum along the imaginary energy axis. Further increase in pump density creates a new
\revision{sideband}
continuum.
 The behavior of the discrete subset is summarized in a phase diagram in the $( n_{\rm p} , \gamma_{\rm cav} )$ parameter space (Fig. \ref{fig:sr1600etcSchematic}). A curve of exceptional points are present which link the Goldstone companion ${\rm G}_1$ to Mollow modes. The overall picture of the low-frequency fluctuation modes that we have found here may also be more generally applicable to other lasers and driven quantum symmetry-broken systems,
e.g.\ \cite{%
ning-haken.90,%
ebata-etal.10,%
hama-etal.13,%
hannibal-etal.18}.

We gratefully acknowledge financial support from NSF under grant number DMR 1839570,
and CPU time at HPC, University of Arizona.

%

\begin{thebibliography}{10}

\bibitem{fan-etal.97pra}
X. Fan, H. Wang, H.~Q. Hou, and B.~E. Hammons, Phys. Rev. A {\bf 56},  3233
  (1997).

\bibitem{cao-etal.97}
H. Cao {\it et~al.}, Phys. Rev. A {\bf 55},  4632   (1997).

\bibitem{gonokami-etal.97}
M. Kuwata-Gonokami {\it et~al.}, Phys. Rev. Lett. {\bf 79},  1341  (1997).

\bibitem{kira-etal.99b}
M. Kira, F. Jahnke, W. Hoyer, and S. Koch, Progress in Quantum Electronics {\bf
  23},  189  (1999).

\bibitem{moskalenko-snoke.00}
S. Moskalenko and D.~W. Snoke, {\em {Bose-Einstein} Condensation of Excitons
  and Biexcitons and Coherent Nonlinear Optics with Excitons} (Cambridge
  University Press, Cambridge, 2000).

\bibitem{Ciuti2000}
C. Ciuti, P. Schwendimann, B. Deveaud, and A. Quattropani, Phys. Rev. B {\bf
  62},  R4825  (2000).

\bibitem{Savvidis2000}
P.~G. Savvidis {\it et~al.}, Phys. Rev. Lett. {\bf 84},  1547  (2000).

\bibitem{kwong-etal.01prl}
N.~H. Kwong {\it et~al.}, Phys. Rev. Lett. {\bf 87},  027402  (2001).

\bibitem{Baumberg2005}
J.~J. Baumberg and P.~G. Lagoudakis, Phys. Stat. Sol. (b) {\bf 242},  2210
  (2005).

\bibitem{Keeling2007}
J. Keeling, F.~M. Marchetti, M.~H. Szymanska, and P.~B. Littlewood, Semicond.
  Sci. Technol. {\bf 22},  R1  (2007).

\bibitem{schumacher-etal.07prb}
S. Schumacher, N.~H. Kwong, and R. Binder, Phys. Rev. B {\bf 76},  245324
  (2007).

\bibitem{bajoni-etal.08}
D. Bajoni {\it et~al.}, Phys. Rev. Lett. {\bf 100},  047401   (2008).

\bibitem{berney-etal.08}
J. Berney, M.~T. Portella-Oberli, and B. Deveaud, Phys. Rev. B {\bf 77},
  121301   (2008).

\bibitem{amo-etal.09}
A. Amo {\it et~al.}, Nature {\bf 457},  291   (2009).

\bibitem{sanvitto-timofeev.12}
D. Sanvitto and V. Timofeev, {\em Exciton Polaritons in Microcavities}
  (Springer, Berlin, 2012).

\bibitem{semkat-etal.09}
D. Semkat {\it et~al.}, Phys. Rev. B {\bf 80},  155201   (2009).

\bibitem{kamide-ogawa.10}
K. Kamide and T. Ogawa, Phys. Rev. Lett. {\bf 105},  056401   (2010).

\bibitem{deng-etal.10}
H. Deng, H. Haug, and Y. Yamamoto, Rev. Mod. Phys. {\bf 82},  1489  (2010).

\bibitem{snoke-littlewood.10}
D. Snoke and P. Littlewood, Physics Today {\bf 63},  42  (2010).

\bibitem{liu-etal.14}
{X. Liu and T. Galfsky and Zh. Sun and F. Xia and E. Lin and Y. Lee and S.
  Kena-Cohen and V. Menon}, Nature Photonics {\bf 9},  30   (2014).

\bibitem{schulze-etal.14}
F. Schulze {\it et~al.}, Phys. Rev. A {\bf 89},  041801(R)  (2014).

\bibitem{menard-etal.14}
J. Menard {\it et~al.}, Nature Comm. {\bf 5},  4648  (2014).

\bibitem{kamandar-etal.14}
D. Kamandar, M. Dignam, M. Steel, and J. Sipe, Phys. Rev. A {\bf 90},  043832
  (2014).

\bibitem{schmutzler-etal.15}
J. Schmutzler {\it et~al.}, Phys. Rev. A {\bf 91},  195308  (2015).

\bibitem{leeuw-etal.16}
A.~W. de~Leeuw {\it et~al.}, Physical Rev. A {\bf 94},  013615  (2016).

\bibitem{hayenga-etal.17}
W. Hayenga and M. Khajavikhan, Light: Science \& Applications {\bf 6},  e17091
  (2017).

\bibitem{kavokin-etal.17}
A. Kavokin, J. Baumberg, G. Malpuech, and F. Laussy, {\em Microcavities}
  (Oxford Science Publications, London, 2017).

\bibitem{bao-etal.19}
W. Bao {\it et~al.}, PNAS {\bf 116},  20274  (2019).

\bibitem{szymanska-etal.06}
M.~H. Szymanska, J. Keeling, and P.~B. Littlewood, Phys. Rev. Lett. {\bf 96},
  230602   (2006).

\bibitem{hu-etal.2019arxiv}
J. Hu {\it et~al.},   (2019).

\bibitem{chou-su.86}
K. Chou and Z. Su, Prog. Theor. Phys. Supp. {\bf 86},  34  (1986).

\bibitem{wouters-carusotto.07a}
M. Wouters and I. Carusotto, Phys. Rev. Lett. {\bf 99},  140402   (2007).

\bibitem{hanai-etal.18}
R. Hanai, P.~B. Littlewood, and Y. Ohashi, Phys. Rev. B {\bf 97},  245302
  (2018).

\bibitem{galitskii-etal.70}
V.~M. Galitskii, S.~P. Goreslavskii, and V.~F. Elesin, Soviet Physics JETP {\bf
  30},  117  (1970).

\bibitem{quochi-etal.98}
F. Quochi {\it et~al.}, Phys. Rev. Lett. {\bf 80},  4733  (1998).

\bibitem{byrnes-etal.10}
T. Byrnes, T. Horikiri, N. Ishida, and Y. Yamamoto, Phys. Rev. Lett. {\bf 105},
   186402   (2010).

\bibitem{horikiri-etal.16}
T. Horikiri {\it et~al.}, Scientific Reports {\bf 6},  25655  (2016).

\bibitem{chow-etal.94}
{W. W. Chow and S. W. Koch and M. Sargent III }, {\em Semiconductor-Laser
  Physics} (Springer, Berlin, 1994).

\bibitem{keller-gallmann}
U. Keller and L. Gallmann, {Ultrafast Laser Physics}.

\bibitem{thedrez-frey.88}
B. Thedrez and R. Frey, Opt. Lett. {\bf 13},  105   (1988).

\bibitem{giorgi-etal.14}
M.~D. Giorgi {\it et~al.}, Phys. Rev. Lett. {\bf 12},  113602  (2014).

\bibitem{horikiri-etal.18}
T. Horikiri {\it et~al.}, J. Phys. Soc. Japan {\bf 87},  094401  (2018).

\bibitem{heiss.04}
W.~D. Heiss, Journal of Physics: Mathematical and General {\bf 37},  2455
  (2004).

\bibitem{dembowski-etal.01}
C. Dembowski {\it et~al.}, Phys. Rev. Lett. {\bf 86},  787   (2001).

\bibitem{liertzer-etal.12}
M. Liertzer {\it et~al.}, Phys. Rev. Lett. {\bf 108},  173901   (2012).

\bibitem{gao-etal.15}
T. Gao {\it et~al.}, Nature {\bf 526},  554   (2015).

\bibitem{hanai-etal.19}
R. Hanai, A. Edelman, Y. Ohashi, and P. Littlewood, Phys. Rev. Lett. {\bf 122},
   185301  (2019).

\bibitem{hanai-littlewood.19}
R. Hanai and P.~B. Littlewood,   (2019).

\bibitem{miri-etal.19}
M.~A. Miri and A. Alu, Science Magazine {\bf 363},  1  (2019).

\bibitem{sakhdari-etal.19}
M. Sakhdari {\it et~al.}, Phys. Rev. Lett. {\bf 123},  193901  (2019).

\bibitem{yamaguchi-etal.13}
M. Yamaguchi {\it et~al.}, Phys. Rev. Lett. {\bf 111},  026404   (2013).

\bibitem{yamaguchi-etal.15}
M. Yamaguchi {\it et~al.}, Phys. Rev. B {\bf 91},  115129   (2015).

\bibitem{ning-haken.90}
C.~Z. Ning and H. Haken, Phys. Rev. B {\bf 41},  3826   (1990).

\bibitem{ebata-etal.10}
S. Ebata {\it et~al.}, Phys. Rev. C {\bf 82},  034306  (2010).

\bibitem{hama-etal.13}
Y. Hama, G. Tsitsishvili, and Z.~F. Ezawa, Prog. Theor. Exp. Phys. {\bf 2013},
  053I01  (2013).

\bibitem{hannibal-etal.18}
S. Hannibal {\it et~al.}, Phys. Rev. A {\bf 97},  013619   (2018).

\end{thebibliography}

\end{document}